\documentclass[conference]{IEEEtran}
\IEEEoverridecommandlockouts
\usepackage{cite}
\usepackage{amsmath,amssymb,amsfonts}
\usepackage{algorithmic}
\usepackage{graphicx}
\usepackage{textcomp}
\usepackage{xcolor}
\usepackage{url}
\def\BibTeX{{\rm B\kern-.05em{\sc i\kern-.025em b}\kern-.08em
    T\kern-.1667em\lower.7ex\hbox{E}\kern-.125emX}}

\usepackage{subcaption}

\begin{document}


\title{Learning to Predict Short-Term Volatility with Order Flow Image Representation}

\author{\IEEEauthorblockN{Artem Lensky}
\IEEEauthorblockA{\textit{School of Engineering and Technology} \\
\textit{University of New South Wales}\\
Canberra, Australia \\
a.lenskiy@unsw.edu.au}
\and
\IEEEauthorblockN{Mingyu Hao}
\IEEEauthorblockA{\textit{School of Computing} \\
\textit{Australian National University}\\
Canberra, Australia \\
mingyu.hao@anu.edu.au}
}

\maketitle



\begin{abstract}
\textit{Introduction}: The paper addresses the challenging problem of predicting the short-term realized volatility of the Bitcoin price using order flow information. The inherent stochastic nature and anti-persistence of price pose difficulties in accurate prediction.

\textit{Methods}: To address this, we propose a method that transforms order flow data over a fixed time interval (snapshots) into images. The order flow includes trade sizes, trade directions, and limit order book, and is mapped into image colour channels. These images are then used to train both a simple 3-layer Convolutional Neural Network (CNN) and more advanced ResNet-18 and ConvMixer, with additionally supplementing them with hand-crafted features. The models are evaluated against classical GARCH, Multilayer Perceptron trained on raw data, and a naive guess method that considers current volatility as a prediction.

\textit{Results}: The experiments are conducted using price data from January 2021 and evaluate model performance in terms of root mean square error (RMSPE). The results show that our order flow representation with a CNN as a predictive model achieves the best performance, with an RMSPE of $0.85\pm1.1$ for the model with aggregated features and $1.0 \pm1.4$ for the model without feature supplementation. ConvMixer with feature supplementation follows closely. In comparison, the RMSPE for the naive guess method was $1.4\pm3.0$. 
\end{abstract}

\maketitle

\section{Introduction}\label{sec1}

A large body of research has been mostly focusing on predicting long-term volatility, ranging from daily to annual prediction\cite{3,7,11}. This makes sense in ensuring that underlying assets, e.g. commodities, will be delivered at the agreed price and the option price will not be too high. Along the classical approaches for volatility prediction e.g. the auto-regressive conditional heteroskedasticity (ARCH)\cite{40}, the generalized auto-regressive conditional heteroskedasticity (GARCH) \cite{41}, the heterogeneous autoregressive model (HAR)\cite{3}, the heterogeneous autoregressive model of realized variant (HAR-RV)\cite{2}\cite{14}, several approaches to predicting long-term volatility using neural networks have been proposed\cite{17}. Ge et al. compared Temporal Convolutional Networks, and Temporal Fusion Transformer among others in predicting the volatility of S\&P500, NASDAQ and several commodities with the help of exogenous inputs such as indices SZSE, BSE SENSEX, FTSE100 and DJIA, exchange rates US-YEN, US-EURO and other fundamentals. The authors demonstrated the superiority of the aforementioned models in predicting the next month's volatility\cite{Ge2023}.  Short-term volatility, on the other hand, is also vitally important for maintaining a "healthy" market structure and in particular providing liquidity. High-frequency trading (HFT) strategies that market-making funds and hedge funds commonly use \cite{12}, employ various extensions of classical autoregressive models. Applying neural networks for short-term volatility prediction is challenging due to the stochastic nature of markets' behavior, implying a lack of repetitive patterns. Hence, information on order distribution and trade dynamics becomes more crucial than at the macroscale (days and years), hence the necessity to analyze different data modalities emerges.

\begin{figure}
     \centering
     \begin{subfigure}{\linewidth}
         \centering
         \includegraphics[width=0.75\linewidth, trim={0.5cm 4.8cm 0.5cm 5cm},clip]{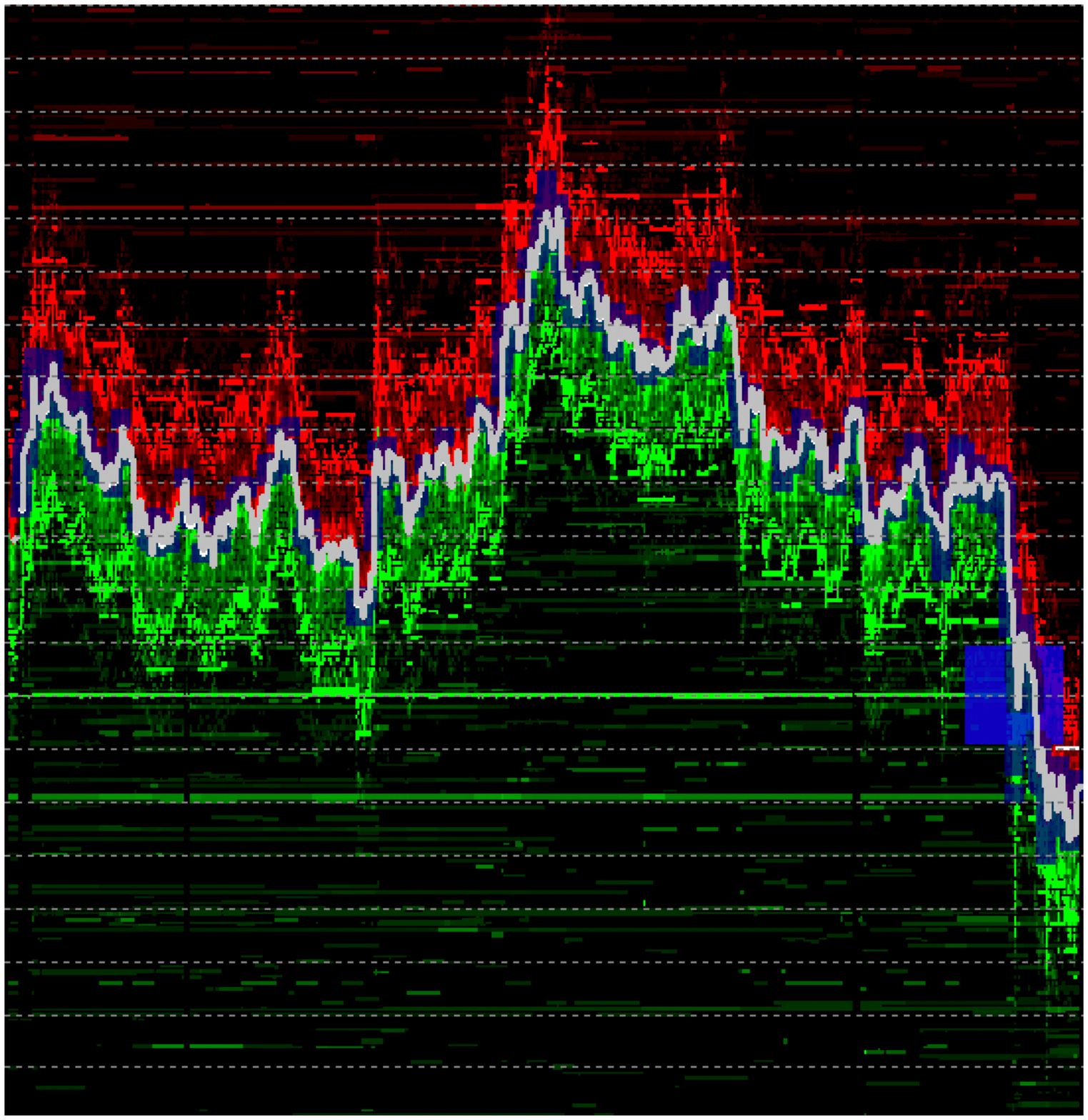}
         \caption{an order flow visualised by a real trading terminal}
         \label{fig:real_terminal}
     \end{subfigure}
     \begin{subfigure}{\linewidth}
         \centering
         \includegraphics[width=0.75\linewidth, trim={4.28cm 22.0cm 12.25cm 3.25cm},clip]{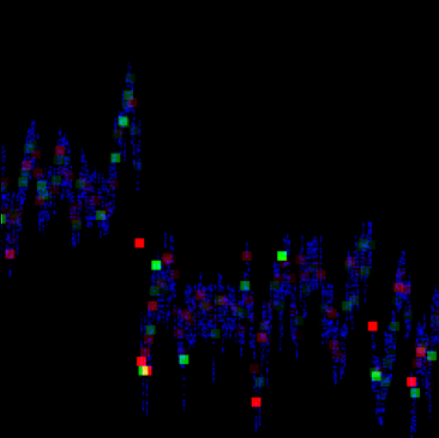}
         \caption{a constructed image representing real order flow}
         \label{fig:simulated_terminal}
     \end{subfigure}
        \caption{(a) The Bid/Ask levels are shown in green/red respectively. The colour intensity is proportional to the size of a limit order. Blue blocks represent large trades, with the size being proportional to the size of a trade e.g. encodes T\&S information. (b) Both sides of the limit orders are shown in blue. The colour intensity is proportional to limit order size. Red and green blocks represent sell and buy trades, with the intensity being proportional to the size of a trade.}
        \label{fig:1}
\end{figure}

The motivation for this paper comes from discretionary traders who employ intraday trading strategies (scalping strategies). This class of traders relies on market order flow to identify support and resistance levels, frequency of trades, and their size distribution to detect market reversals and breakouts. Figure \ref{fig:1}(a) illustrates an example of the order flow that visualizes the Bid and Ask sides of the limit order book and large trades. The figure illustrates an event when a large Bid limit order sitting in the book is filled by a large market sell order (a large blue square). This large selling indicates a sell pressure and hence a downtrend might be anticipated by the trader. The paper tests the hypothesis of whether a relatively simplistic Convolutional Neural Network (CNN) can learn to capture such events and predict price volatility short-term as well as compare it to more sophisticated models such as ConvMixer and ResNet18. The paper also illustrates that the proposed approach allows for interpretability.

\section{Terminology} \label{sec: formulating tAsk}

\subsection{Realised Volatility}
The paper aims at predicting the realized volatility. As defined by \cite{22} and \cite{33}, the realized volatility measures the change in price by taking the square root of the sum of the squared returns. Defining the log return for time $t$ and $t+\tau$ as $r_{t, \tau}=P(t+\tau)-P(t) \label{eq2}$, where $P(t)=\texttt{log}(p_t)$ and $p_t$ is price at time $t$, the realized variance $R$ is calculated as $R_{t+\tau} = \sum^n_{i=0} r^2_{\tau_i, \tau_{i+1}}$
where $t <= \tau_0 < \tau_1 < ... < \tau_n <=t+\tau$ are timestamps of sample points from the target period.  Finally, the realized volatility is simply the square root of the realized variance $RV = \sqrt{R}$. An accurate prediction of future volatility will help traders anticipate potential risks and opportunities in the future. The objective is to predict $RV$ on $(t, t+\tau)$, where $\tau$ is the forecast horizon. 

\subsection{Limit order book}
The Limit Order Book (LOB) provides vital information on short-term asset behavior \cite{21}\cite{27}. The difference in price and order sizes at the Bid and Ask sides (order book imbalance) carry information that trading agents might use to make trading decisions. Moreover, the short-term price dynamics might depend on the frequency of trades, trade sizes, and the correlation in trade sides, and hence can contribute to explaining short-term volatility. Therefore, a hypothesis is tested that the Time \& Sales (T\&S) and the LOB provide vital information for improving short-term volatility forecasting. 

\subsection{Order Flow}
The evolution of the LOB over time is called order flow and is a tool used by traders and trading algorithms. Formally, an order $O$ consists of three parts: the executed time $t_O$ (or current time for listed but not yet executed orders, limit orders), the order price $p_O$, and order volume or size $V_O$, resulting in a tuple $O := (t_O, p_O, V_O)$.

The T\&S records represent executed/market orders, while the LOB stores the top limit orders, i.e., the price levels at which a buyer or a seller is willing to buy or sell correspondingly. The levels are sorted by price for both the Bid and Ask sides. Then, for a given time $t$, the order flow $F$ refers to all orders, listed in the LOB or executed (LOB and T\&S) between $(t-\Delta T, t)$, that is $O \in F_{t-\Delta T, t} \iff t-\Delta T \leq t_) \leq t$, where $\Delta T$ is the input size of the time series. An example of order flow superimposed with large trades is shown in Figure \ref{fig:1}.

\section{Methodology}\label{pipeline}

\subsection{Data} \label{dataset}
The perpetual futures contract (PFC) of BTC / USDT provided by the Binance public historical market data set\cite{36} was used for experimentation. BTC/USDT represents the value of Bitcoin in tether dollars (USDT).  According to Binance, among all the PFCs, the BTC/USDT PFC makes the majority of the daily trading volume. The data contains tick-level trades and 20 levels of limit order book on the Bid and Ask sides. Generally, the price of the contract closely follows the price at the start of the Bitcoin market due to the ``funding fee'' mechanism. The funding fee is a varying payment between long/short position traders depending on the difference between future contract price (mark price) and spot price that ensures that the mark price follows closely with the spot price. For example, when the mark price is higher than the spot price, traders at long positions pay extra fees to traders at short positions.  Therefore, the volatility of perpetual futures contracts also mostly replicates that of the Bitcoin spot price. It is important to note that BTC/USDT can be traded with a leverage of up to 125, which could and usually results in slightly higher volatility due to potentially trigger stop cascades \cite{Liston_2023}. It is also worth noting that besides speculative trading, PFC is widely used for hedging against volatility risk in the cryptocurrency market.

\begin{table*}[t]
\caption{Information presented as part of the dataset}
  \centering
    \begin{tabular}{ p{0.08\textwidth}p{0.32\textwidth}|p{0.08\textwidth}p{0.35\textwidth}}
    \hline
    \multicolumn{2}{c}{Order Book} & \multicolumn{2}{c}{T\&S} \\
    \hline
    Ask price   & Ask price from level 1 to level 20 & price & weighted average executed trade price in last second\\
    Bid price  & Bid price from level 1 to level 20   & order count & number of executed trade orders in last second\\
    Ask size   & volume per Ask price from level 1 to level 20  & size & total volume of executed trades in the last second\\
    Bid size   & volume per Bid price from level 1 to level 20 & is\_buyer &  whether the buyer is the market maker \\
    \hline
    \end{tabular}
    \label{tab1}
\end{table*}

\subsection{Image Encoding}
In order to simulate the market information presented to the trader (Figure \ref{fig:1}(a)), an encoding heuristic that stores LOB and T\&S information as a single image is developed. This also provides the means for employing either pre-trained or designed-from-scratch CNNs. It is important to note, that the idea of encoding market data into images is not new, for example, Ge et al. trained LSTM-CNN to predict monthly volatility\cite{Ge2023} using images constructed from price data by employing Gramian Angular Fields and Markov Transition Fields \cite{wang_encoding_2015}\cite{kristjanpoller_gold_2015}, however, these encodings are not intuitive and rely on mathematical tricks in the hope of improving the predictability, while our encoding simulates the representation used by professional traders.

A resolution of $m\times n$ pixels with $c$ channels is chosen, where each channel encodes a specific piece of information. A unit time interval $t$ and a unit value interval $v$ are further defined. Let each column of pixels represents information in $(t_i, t_{i+1})$ of input data, and the rows represent order information in $(v_j, v_{j+1})$. The pixel value, ranging from 0 to 255, represents the normalized volume of limit orders at the corresponding row (value interval) and column (time interval). 
\begin{equation}
\texttt{pixel}_{t_i, v_j} = (V_o) \iff  [t_O \in (t_i, t_{i+1})] \land [p_O \in (v_j, v_{j+1})]
\end{equation}
Then for each channel $c_i$, the image encodes any order settled in the time interval $(t_0, t_0+n \cdot t)$ with order value in the range of $(v_0, v_0+m \cdot v)$, where $t_0$ is the starting time of the input data, and $v_0$ is the minimum price which is captured for image encoding. Formally, for an image $\texttt{IMG}$, the $c_i$ channel encodes all orders in the order flow $F$, if the order is within the range of the image:
\begin{equation}
O \in \texttt{IMG}_{c_i} \iff [O_t \in (t_0, t_0+n \cdot t)] \land [p_O \in (v_0, v_0 + m \cdot v)]
\end{equation}

To avoid confusion among multiple data modalities, each channel should only encode one piece of information. Intuitively, such an image is a 'snapshot' of order flow similar to Figure \ref{fig:1}(a), where pixels represent the order's time and value, while colours represent different modalities of the order flow. The value of $t$ and $v$ determines the 'resolution' of the snapshot. The size of the image determines the range of encoded data. Further, $\epsilon$ is defined as the time interval between each consecutive encoded image. When $\epsilon < t$, the encoded images overlap with each other, and when $\epsilon > t$, the image set is a sub-sampling of the original order flow. 

To align with the input size of common pre-trained CNN models such as ResNet\cite{19} or ConvMixer\cite{trockman2022patches}, the image resolution is set to $240 \times 240$, with $t$ measured in seconds and $v$ in USDT. Hence each image contains information on the order flow for 4 minutes and the range is 240 tether dollars. The Red and Green channels encode market orders (trades). To amplify the effect of a market order, each trading pixel is padded to form a square. This will result in overlapping squares when several trades occur at about the same time, and the value of overlapped pixels is the sum of all these trades. Such implementation allows us to highlight regions with more active trading activities, which may lead to changes in volatility. Buy and sell orders are encoded by green and red channels correspondingly. 

The Blue channel encodes the LOB. The limit orders are unfortunately limited to the top 20 levels. For instance, in the case when the price goes down, the top of the LOB becomes occupied by new orders with lower prices, and the original top orders disappear from the LOB. However, since the old orders are not filled, and as long as the trader still has the intent to trade, an assumption is made that the order will be kept in the LOB at the same price level and with the same size until new information indicates that the order has changed, or the order is filled. In practice, agents may change their position by actively tracking the price as demonstrated in Figure \ref{fig:1}(a). Figure \ref{fig:1}(b) illustrates an example of encoded market data following the aforementioned procedure.


\subsection{The pipeline}
\label{cnn}
Based on the above trading information representation, the encoded image is generally suitable for any conventional 2D-CNN model. Several models with different structures are investigated to validate the performance using the proposed encoding method and further search for the most appropriate model. 

\begin{figure*}[t!]
\centerline{\includegraphics[width=1\linewidth, trim={0.5cm 3.5cm 3.2cm 0.5cm},clip]{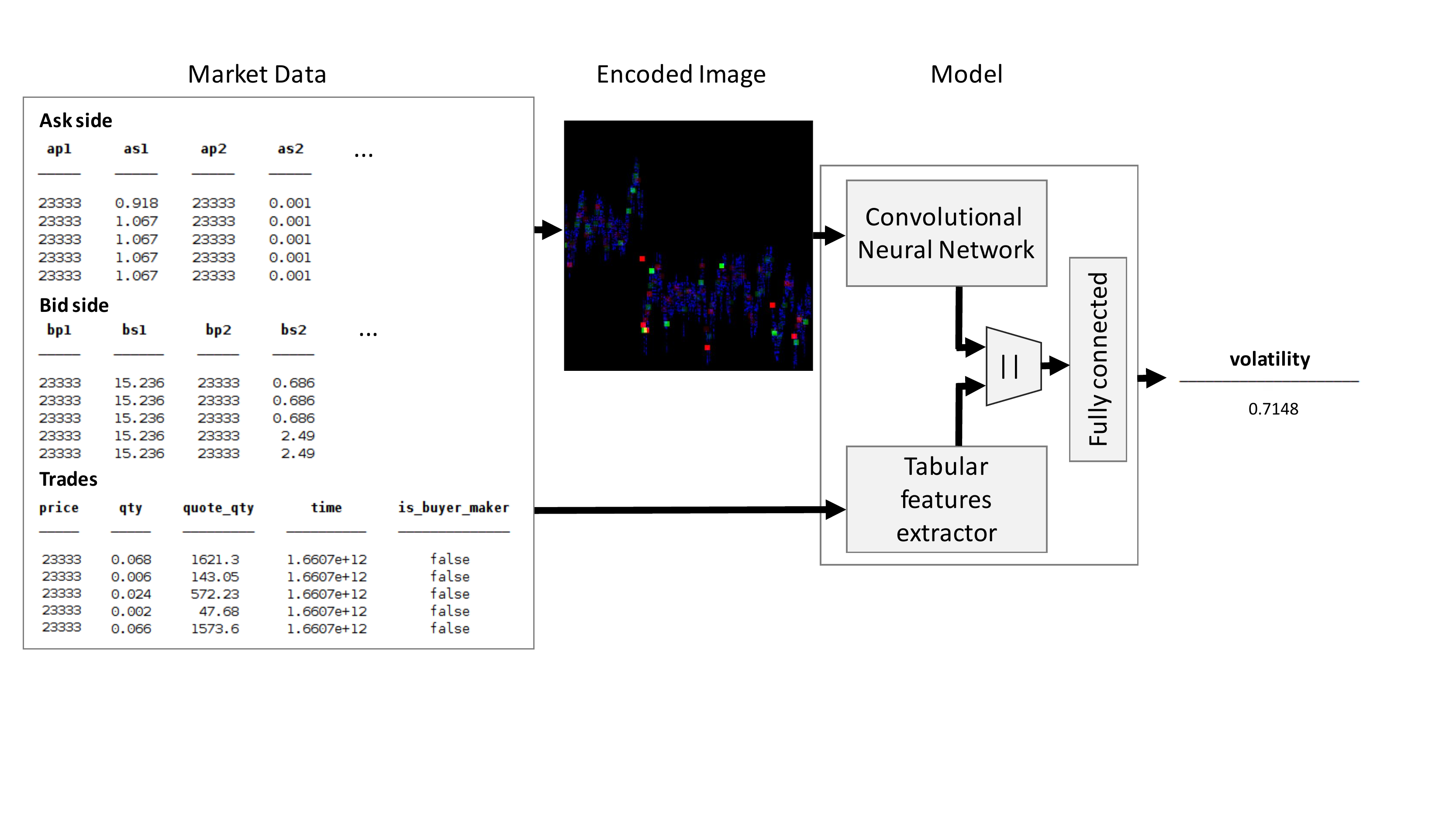}}
\caption{Order flow is passed to the tabular feature extractor and to the image encoder. The encoded images are then passed to the CNN model. The outputs of both parts are concatenated and fed into the regression layer(fully connected layer).}
\label{fig:2}
\end{figure*}

The main model under testing consists of two parts: (a) the CNN, either a pre-trained one or trained from scratch and (b) a tabular (manually crafted) feature extractor as shown in Figure \ref{fig:2}. The last layer of the network is replaced with a flattening layer, to produce a 1D vector. The features extracted from encoded images by the CNN model are combined with externally extracted features. Based on \cite{27} and \cite{34}, a set of features is selected consisting of 393 aggregated features. Finally, a regression layer maps the aggregated feature vector to a predicted volatility value. 

\subsection{Models}
The full list of the tested models with insights into their structural parameters is outlined next.

\textit{Naive Guess} simply takes realized volatility of the last minute and uses it as a prediction.

\textit{GARCH} is a classical univariate model that uses past squared returns to predict volatility. The model requires two parameters order of autoregression ($p$), and order of moving average ($q$). A search over these parameters is performed by training the model with a relatively small subset. The model with $p=1, q=1$ had the best performance. 

\textit{Multilayer Perceptron} (MLP), is a fully connected network trained on the aggregated feature set described in \ref{dataset}.

\textit{CNNPred-2D} is based on \cite{25} and takes multivariate input of dimension $85 \times 240$. Typically the model has a first-layer kernel size of 1 to aggregate all features at each timestamp into one latent feature. The last fully connected layer is modified for regression. 

\textit{CNN-1D} takes raw order flow as input i.e. each second is represented by an 85-dimensional vector. The model is typically with 6 convolution layers with filter size $5$, followed by pooling layers and affine regression layers. 

\textit{LSTM-1D} takes raw order flow as input (similarly to the CNN-1D). The model uses three LSTM layers followed by an affine regression layer.

\textit{ResNet-18}, is a pretrained benchmark CNN-type of model. The model has 18 residual blocks and each residual block consists of 2 convolution layers with kernel size $3 \times 3$ and batch normalizations. The last layer is replaced by a regression layer.

\begin{figure*}[t!]
    \centerline{\includegraphics[width=1\linewidth, trim={3.25cm 4.5cm 12.5cm 4.5cm},clip]{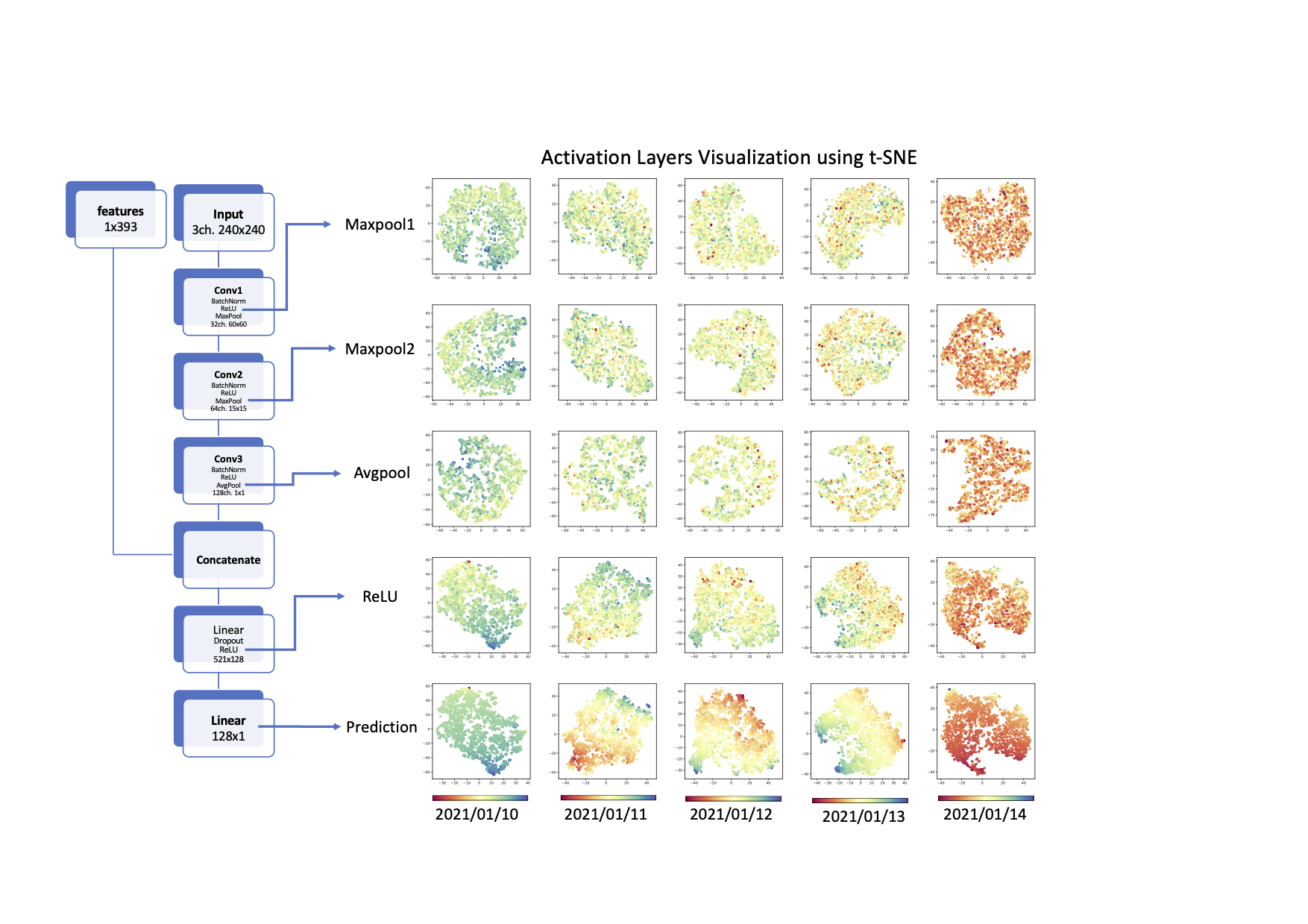}}
    \caption{The left-hand diagram shows the structure of the base CNN model, and the on the left are the weights of corresponding layers visualise with t-SNE  at $\texttt{0:05}, \texttt{20}^{th}, \texttt{June}, \texttt{2022}$ to $\texttt{23:50}, \texttt{24}^{th}, \texttt{June}, \texttt{2022}$. The colour band corresponds to the volatility range of the day. Notice that because the volatility is normalized separately for each day, there is no absolute correlation between samples with the same colour across days. }
    \label{fig:10}
\end{figure*}

\textit{Naive-CNN} is used to evaluate solely the CNN part without the aggregated feature vectors. The model has 3 convolution layers. Each layer is followed by batch normalization, pooling and ReLU activation. The last layer is a regression layer mapping 128 latent features into one output. The model configurations are shown in Figure \ref{fig:10}.

\textit{ConvMixer-Aggr} is similar to CNN-Aggr but with CNN being replaced by the pretrained state-of-the-art ConvMixer\cite{trockman2022patches}. The model starts with a ``patch layer'' with patch size 2 followed by batch normalization and GeLU activation \cite{hendrycks2016gaussian}. Following the patch layer are 2 residual blocks where each residual block consists of one convolution layer with kernel size 5 and the other convolution layer with kernel size 1.

\section{Results} \label{settings}
Prior to image normalization, pixels representing the trade size were clipped to the 99 percentile, in order to remove outliers, that otherwise suppress the intensities of all other pixels. The prediction target was the realized volatility in the next 60 seconds. Table \ref{tab1} outlines the information presented at every second in an input window. The images are generated in a walk-forward approach. The $\epsilon$ is set to 10 seconds, so each pair of neighbouring images have 230 seconds of overlaps. This results in 8616 images for each trading day. To avoid information leaks when training the models, the images are fed in time order. The dataset is split into the train, validation, and test sets with a size ratio of 3:1:1, where the timestamps of the test set are strictly after the training set to prevent information leaks. The performance of the aforementioned model was evaluated using the average Root Mean Square Percentage Error(RMSPE) $\rho = \sqrt{\frac{1}{N}\sum_{k=1}^N\left( \frac{\hat{y}_k-y_k}{y_k}\right)^2}$ and compared to the aforementioned models:
\begin{table*}[t]
\setlength{\arrayrulewidth}{0.3mm}
\setlength{\tabcolsep}{7pt}
\renewcommand{\arraystretch}{1.0}
  \centering
      \caption{The mean and standard deviation of RMSPE computed for validation and test sets of the models trained on daily data\\ from $1^{th} Jan, 2021$ to $30^{th} Jan, 2021$}
    \begin{tabular}{p{0.2\textwidth}p{0.25\textwidth}p{0.25\textwidth}}
    \hline
    Models & validation set & test set\\
\hline
Naive guess & $0.414 \pm 0.128$ & $1.398 \pm 2.992$\\
GARCH       & $0.889 \pm 0.603$ &  - \\
MLP         & $0.582 \pm 0.249$ & $1.046 \pm 1.099$\\
CNNPred-2D  & $0.365 \pm 0.084$ & $1.079 \pm 1.422$\\
CNN-1D      & $0.579 \pm 0.200$ & $1.288 \pm 1.435$\\
LSTM-1D     & $0.346 \pm 0.064$ & $1.116 \pm 1.709$\\
ResNet-18   & $0.350 \pm 0.056$ & $1.225 \pm 2.005$\\
Naive-CNN   & $0.355 \pm 0.060$ & $1.005 \pm 1.407$\\
CNN-Aggr    & $\mathbf{0.309} \pm 0.071$ & $\mathbf{0.851} \pm 1.110$\\
ConvMixer-Aggr & $0.519 \pm 0.184$ & $0.994 \pm 1.057$ \\
\hline
    \end{tabular}
    \label{tab2}
\end{table*}

The Root Mean Square Percentage Error (RMSPE) obtained for the range of $1^{th} Jan, 2021$ to $30^{th} Jan, 2021$ are presented in table \ref{tab2}. On average the proposed CNN-Aggr model outperforms all other models, with the best average score. It also has the lowest standard deviation indicating a robust overall performance. More heavier models such as ResNet-18 and ConvMixer-Aggr did not show the best performance as one might have expected, this might be due to the fact that these models were pretrained on ImageNet consisting of real object images that are very different from the synthetic order flow images.  Surprisingly, the naive guess can sometimes provide fairly accurate forecasting, especially when the price is stable across the day i.e. the corresponding minute volatility does not change too much. However, the standard deviation of naive guesses is much higher, implying less consistent prediction accuracy. Comparing the Naive-CNN and the CNN-Aggr models, the aggregation of additional features does improve the accuracy of prediction. To verify that the order book dynamics actually provide valuable market insights, the MLP model is trained solely on the hand-crafted features. This model performed better than Naive guess but worse than Naive-CNN, implying that the market dynamics represented in the form of images carry important information that is not captured by the additional features and vice versa, which explains why the CNN-Aggr model that can be seen as a combination of the MLP and naive CNN models outperformed each of these models alone. 


Further, CNN-LSTM models are experimented with. The models use embeddings extracted from CNN as LSTM input. The embedding output of consecutive images from CNN is fed into the LSTM model as a time series. However, the models failed to further improve the prediction, regardless of whether the CNN-Aggr model or pretrained ResNet was used, this could be due to images being encoded and normalized individually, which results in breaking the correlation between consecutive images.  The stochastic nature of the price behaviour makes the accumulated prediction residual error large.

\begin{figure*}[htbp!]
    \centerline{\includegraphics[width=1\linewidth, trim={6.0cm 12.5cm 16cm 12.5cm},clip]{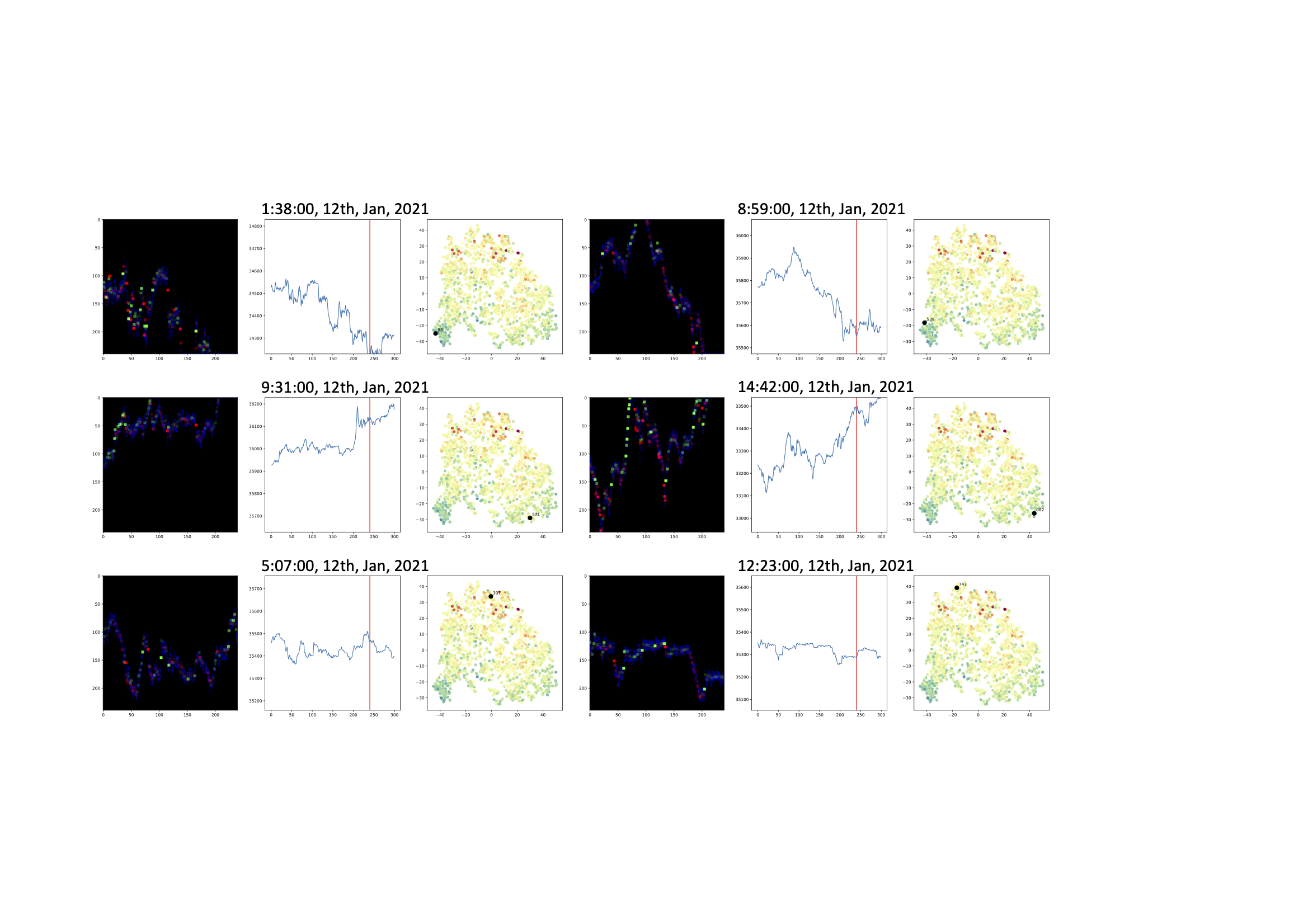}}
    \caption{t-SNE visualization of final ReLU layer on $21^{th}, June, 2022$. Six samples represent different market conditions. The left image is the input; the middle graph depicts the market price at the same time period (including both the input and target period on the left and right side of the red line); the right graph shows the position of the sample (black dot) in the t-SNE plot of last activation layer.}
    \label{fig:11}
\end{figure*}

\section{Interpretability} \label{sec: discussions}

To interpret the behaviour of the models, the input data are passed through the trained model and output embedding vectors from each activation layer are recorded. Four activation layers are chosen for investigation as shown in Figure \ref{fig:10}. Notice that the embedding vectors are of high dimension (e.g output of the Maxpool1 layer consists of 115200 values after being flattened), in order to visualize the distribution of these high-dimensional embeddings t-Stochastic Neighbor Embedding (t-SNE) \cite{39} is employed. t-SNE struggles in mapping from very high dimensional space and hence the input space is  first processed by Principle Component Analysis (PCA) reducing the dimensionality to 40. Figure \ref{fig:10} shows the sample visualization results layer by layer for CNN-Aggr models trained on 5 different periods from the latest available daily order flows. Each point represents an input sample, with colours representing the normalized target volatility. At the first activation layer, the distribution of embedding vectors does not show any obvious patterns. As the input data propagate through layers, samples with high volatilities (green) and low volatilities (red) tend to separate. At the last activation layer, there is a clear gradual change in colour along the axes. High volatile samples (green) are split into two corners. Further analysis shows that the two corners correspond to an upward trend and a downward trend of input data. For the model trained on $21^{th}, June, 2022$, samples with clear downward trends and higher volatilities are more likely to be mapped to the top left of the plot, while samples with highly volatile upward trends are mapped to the bottom left corner (Figure \ref{fig:11}). Samples representing trades within a trading channel i.e. sideways price movement and those associated with non-obvious trends are mapped to the right part of the plot. This corresponds to the prediction distribution shown in the last line in Figure \ref{fig:10}. Generally, the model gives a higher volatility prediction for samples on the left side than on the right. Intuitively, this means a highly volatile market is likely to have high volatility in the future as well. 

The choice of $v_0$ and $v$ is quite important but challenging, that is due to the variability of the stock price range. Ideally, the range should capture as many orders as possible. When $v$ is too large, the order flows are squeezed into a narrow band and hence it becomes troublesome to extract useful information. On the contrary, when $v$ is too small, the orders outside the value range become hidden from the model. By trial and error, $v$ was set to $1$, while $v_0=O_{p_{0}}-m/2$. It would be interesting to further explore the effect of different value ranges.

\section{Conclusion and Future works}
An encoding method that maps order flow of market and limit orders into images has been proposed. Such image representation provides useful insight into future volatility by better capturing the context of price events. Several models have been trained on these data. Experiments show that a simple CNN model complemented by the aggregated features achieves the best prediction accuracy. The proposed pipeline can be easily generalized to predict the volatility of other asset classes as long as the top of the order book and the time and sales are provided. 

An important difference between the images representing order flows and images representing real-world objects is the temporal property of the former ones. To account for that, an attention-based temporal convolutional neural network \cite{16} is probably a better choice. As we have shown the configuration of the encoded images such as value range and unit value of pixels significantly affect the encoding quality. It is worth further investigating various network configurations in order to validate the robustness of the market data representation as well as the models to further improve the prediction accuracy.

\bibliographystyle{iet}
\bibliography{ref-add}

\begin{thebibliography}{10}

\bibitem{3}
Luo, J., Chen, L.: `Realized volatility forecast with the bayesian random compressed multivariate har model', \emph{International Journal of Forecasting},  2020, \textbf{36}, (3), pp.~781--799

\bibitem{7}
Jorion, P.: `Predicting volatility in the foreign exchange market', \emph{The Journal of Finance},  1995, \textbf{50}, (2), pp.~507--528

\bibitem{11}
Idrees, S.M., Alam, M.A., Agarwal, P.: `A prediction approach for stock market volatility based on time series data', \emph{IEEE Access},  2019, \textbf{7}, pp.~17287--17298

\bibitem{40}
Engle, R.F.: `Autoregressive conditional heteroscedasticity with estimates of the variance of united kingdom inflation', \emph{Econometrica: Journal of the econometric society},  1982, pp.~ 987--1007

\bibitem{41}
Hansen, P.R., Lunde, A.: `A forecast comparison of volatility models: does anything beat a garch (1, 1)?', \emph{Journal of applied econometrics},  2005, \textbf{20}, (7), pp.~873--889

\bibitem{2}
Barun{\'\i}k, J., K{\v{r}}ehl{\'\i}k, T.: `Combining high frequency data with non-linear models for forecasting energy market volatility', \emph{Expert Systems with Applications},  2016, \textbf{55}, pp.~222--242

\bibitem{14}
Corsi, F.: `A simple approximate long-memory model of realized volatility', \emph{Journal of Financial Econometrics},  2009, \textbf{7}, (2), pp.~174--196

\bibitem{17}
Ge, W., Lalbakhsh, P., Isai, L., et~al.: `Neural network--based financial volatility forecasting: A systematic review', \emph{ACM Computing Surveys (CSUR)},  2022, \textbf{55}, (1), pp.~1--30

\bibitem{Ge2023}
Ge, W., Lalbakhsh, P., Isai, L., et~al.: `Comparing deep learning models for the task of volatility prediction using multivariate data', \emph{arXiv preprint arXiv:230612446},  2023,

\bibitem{12}
Zhang, F.: `High-frequency trading, stock volatility, and price discovery', \emph{Available at SSRN 1691679},  2010,

\bibitem{22}
Andersen, T.G., Bollerslev, T.: `Answering the skeptics: Yes, standard volatility models do provide accurate forecasts', \emph{International economic review},  1998, pp.~ 885--905

\bibitem{33}
McAleer, M., Medeiros, M.C.: `Realized volatility: A review', \emph{Econometric reviews},  2008, \textbf{27}, (1-3), pp.~10--45

\bibitem{21}
Qureshi, F.I.: `Investigating limit order book characteristics for short term price prediction: a machine learning approach', \emph{arXiv preprint arXiv:190110534},  2018,

\bibitem{27}
Chen, Q., Robert, C.Y.: `Multivariate realized volatility forecasting with graph neural network', \emph{arXiv preprint arXiv:211209015},  2021,

\bibitem{36}
Binance. `Binance public data'. (,  2022.
\newblock Available from: \url{https://github.com/binance/binance-public-data}

\bibitem{Liston_2023}
Liston, P., Gretton, C., Lensky, A.
\newblock `Modeling the effect of cascading stop-losses and its impact on price dynamics'.
\newblock In: AI for Financial Institutions. (AAAI Bridge 2023 FinST,  2023.

\bibitem{wang_encoding_2015}
Wang, Z., Oates, T.
\newblock `Encoding {Time} {Series} as {Images} for {Visual} {Inspection} and {Classification} {Using} {Tiled} {Convolutional} {Neural} {Networks}'.
\newblock In: Workshops at the {Twenty}-{Ninth} {AAAI} {Conference} on {Artificial} {Intelligence}. (,  2015. p.~8

\bibitem{kristjanpoller_gold_2015}
Kristjanpoller, W., Minutolo, M.C.: `Gold price volatility: {A} forecasting approach using the {Artificial} {Neural} {Network}–{GARCH} model', \emph{Expert Systems with Applications},  2015, \textbf{42}, (20), pp.~7245--7251.
\newblock Available from: \url{http://www.sciencedirect.com/science/article/pii/S0957417415003000}

\bibitem{19}
He, K., Zhang, X., Ren, S., et~al.
\newblock `Deep residual learning for image recognition'.
\newblock In: Proceedings of the IEEE Conference on Computer Vision and Pattern Recognition (CVPR). (,  2016.

\bibitem{trockman2022patches}
Trockman, A., Kolter, J.Z.: `Patches are all you need?', \emph{arXiv preprint arXiv:220109792},  2022,

\bibitem{34}
Nyanpn. `1st place (public 2nd place) solution at kaggle optiver realized volatility prediction'. (Kaggle,  2022

\bibitem{25}
Hoseinzade, E., Haratizadeh, S.: `Cnnpred: Cnn-based stock market prediction using several data sources', \emph{arXiv preprint arXiv:181008923},  2018,

\bibitem{hendrycks2016gaussian}
Hendrycks, D., Gimpel, K.: `Gaussian error linear units (gelus)', \emph{arXiv preprint arXiv:160608415},  2016,

\bibitem{39}
Van~der Maaten, L., Hinton, G.: `Visualizing data using t-sne.', \emph{Journal of machine learning research},  2008, \textbf{9}, (11)

\bibitem{16}
Liu, R.: `Attention Based Temporal Convolutional Neural Network for Real-Time 3D Human Pose Reconstruction'.
\newblock (University of Dayton,  2019)

\end{thebibliography}

\end{document}